\documentclass[letterpaper,11pt]{report}

\usepackage{fullpage}
\usepackage{verbatim}
\usepackage{cite}
\usepackage{setspace}
\usepackage{fancyhdr}
\usepackage{titlesec}

\usepackage[small]{caption}
\usepackage{graphics}
\usepackage{color}

\usepackage{hyperref}

\usepackage[dvips]{graphicx}
\titleformat{\chapter}{\normalfont\huge}{\thechapter.}{20pt}{\huge}

\def\title{Mytitle}

\begin{document}

\begin{titlepage}
   \vspace*{\stretch{1.0}}
   \begin{center}
      \Huge\textbf{Crowd Sourced Data Analysis: Mapping of Programming Concepts to Syntactical Patterns}\\[1\baselineskip]
      \large\textit{Deepak Thukral (deepak14036@iiitd.ac.in) \& Darvesh Punia (darvesh14034@iiitd.ac.in)}
   \end{center}
   \vspace*{\stretch{2.0}}
\end{titlepage}

\include{btech_thesis_cover}

%\newpage
%
%\pagestyle{empty}
%\vspace*{7.1in} 

\pagebreak

\begin{abstract}

Since programming concepts do not match their syntactic representations, code search is a very tedious task. For instance in Java or C, array doesn’t match [], so using “array” as a query, one cannot find what they are looking for. Often developers have to search code whether to understand any code, or to reuse some part of that code, or just to read it, without natural language searching, developers have to often scroll back and forth or use variable names as their queries. In our work, we have used Stackoverflow (SO) question and answers to make a mapping of programming concepts with their respective natural language keywords, and then tag these natural language terms to every line of code, which can further we used in searching using natural language keywords.

\vspace{2in}
\end{abstract}

\newpage

\tableofcontents

%\newpage

%\newpage

% \newpage
\mbox{}

%\doublespacing

\chapter{Introduction}\label{chapter:introduction}
\pagenumbering{arabic}
\setcounter{page}{1}
\onehalfspacing
% \cite{Perugini:2007:SOI:1240624.1240770}

In many community-based information web sites, such as Stack Overflow, users contribute content in the form of questions and answers, which allows others to learn through the contributions of the community.  \\
\textbf{In our project we intend to use Stack Overflow as a tool to improve source code search.}\\
We focus on questions related to \textit{\textbf{Java}} for research purposes.
We consulted the paper "Ranking Crowd Knowledge to Assist Software Development" which categorizes all the questions in 4 types. We realized that categorization of questions could be good step in solving our problem.\\
The questions were categorized into broadly 4 categories: \textbf{Debug}, \textbf{How To Do It}, \textbf{Seeking Different Solution} and \textbf{Need To Know/Conceptual.} \\
Searching in code is often done by developers, where looking through thousands of lines of code to find things is not only time consuming but also tiring. Developers always need to search for code fragments when they write code, or when they want to debug, look up code fragments to reuse or try to understand somebody else's code.
In our work we create a mapping of programming concepts with their syntactical patters using StackOverflow question and answers. We created it using three major steps: \textbf{Entity Discovery}: Parts of Speech (POS) technique is used to discover more entities, \textbf{Mapping Creation}: In this we extract syntactical patters matching their programming concepts, \textbf{Entity Linking}: Finally, each line of source code is annotated with it's associated natural language terms using the mapping created, which further allows keyword based searching.

\chapter{Literature Survey}\label{chapter:literature survey}
\onehalfspacing

\textbf{Word2Vec}\\
Word2Vec is a two layered neural that is trained to represent a word. For input it requires a huge corpus of text. And its output is a vector for the words present in the input corpus. The word vectors even carry the contextual information for each of the word. One typical example of how fantastic the technique can work is that in vector representation "King" - "Man" + "Woman" = "Queen". \\
 
\textbf{Doc2Vec}\\
Doc2vec adapts word2vec to a paragraph or a document. It represents a document by a n-dimensional vector. Each paragraph is often represented by a tag and this tag can be used to find similar documents by their tag. It is often used for document level matching.\\

\textbf{Topic Modelling}\\
Topic modelling is an unsupervised algorithm used in text mining that can extract topics from a given text. Topic modelling is typically done by using LDA. It is able to generate topics that may be hidden in the intricacies of the article.\\

\textbf{Latent Dirichlet Allocation (LDA)}\\
LDA assumes that there is some distribution of topics in the document that generate any words and hence particular document. Initially, any random topic distribution is assumed. Then, for each document the mixture of topics is found. Then, assuming the underlying distribution of topics, the probability of getting that document is generated. This is done over several iterations ultimately getting a stable generative model. \\

\textbf{Term frequency}\\
There are several ways to incorporate the frequency of any word in its weight. A typical way is to simply use the frequency count of that word as its term frequency.\\

\textbf{Inverse document frequency}\\
This term penalises the weight if it appears in several document. It measures how important a given term could be in the entire document depending on how rare it is. Mathematically, it is written as 
 \begin{equation}
     idf(w,D) = log \frac{N}{|\{ d : w \in d,  d \in D \}|}\
 \end{equation}
 where N is the total number of documents and D represts all the documents.\\

\textbf{Term frequency Inverse document frequency}\\
Tf-idf is finally computed simply as the product of tf and idf for any given word.

\chapter{Dataset and Preprocessing}\label{chapter:dataset and preprocessing}
\onehalfspacing

We chose java as the topic for research purpose.	 Although, at times the questions were further rewritten to test various techniques. \\

We adopted the following preprocessing steps
\begin{enumerate}
    \item \textbf{Lemmatization}: The goal of lemmatization is to reduce inflectional forms and sometimes derivationally related forms of a word to a common base form.
    \item \textbf{Tokenizing}: We tokenized the documents to its words.
    \item \textbf{Stop Words Removal}: Typical words like “the”, “is” do not carry much meaning and are called stop words. We removed stop words in some of the techniques.
\end{enumerate}

Entity Discovery: We started off with 20 manually taken entities that belonged to "Java" language from Tutorialspoint, and then ran it over questions obtained from posts having Tag as "java".

Entity-Profile Construction: In this we require SO posts, which are have tag attribute of specific language (in our case we have taken the language Java), and has at least one code snippet.

For this we had taken 7500000 posts( 15 files of 500000 posts each ), and after processing we obtained 110000 posts, which had tag of "Java" and contained at least one code snipper per post.

\chapter{Categorization of questions}\label{chapter:Categorization of questions}

We started with the paper \textbf{Ranking Crowd Knowledge to Assist Software Development’} which discusses the categorization of questions into different types and finds the characteristic/attributes of an effective answers. It also focuses on finding the kind of answers or code examples which help developers and maintainers solve their problems and how are they different from not-so-helpful examples.\\
\begin{enumerate}
    \item \textbf{Debug/Corrective:} This class of questions generally deals with correcting the errors in the program such as Runtime errors, exceptions or a broken code..\\
    \item \textbf{Need to Know:} These set of questions are generally conceptual and seek some knowledge about the topic, technology, language or any API etc.\\
    \item \textbf{How-to-do-it:} This category deals with questions asking the approach to a given problem and may be implementation details as well.\\
    \item \textbf{Seeking different solution:} The questions seeks a different/better approach or solution for an already solved problem.\\
\end{enumerate}

\section{Topic Modeling using LDA}

LDA was applied on the questions of StackOverflow to divide the text into topics. Top few words of each topic were then examined as shown in the results. The main parameter to the model was number of topics. To get more clear results we mixed 200 articles of \url{https://en.wikipedia.org/wiki/Wikipedia:WikiProject_Java/List_of_articles}. We ensured that noise level is low so that there is not much difference in results. to  We kept no of topics equal to 4 initially as discussed in the paper. 

\subsection {Results}

Topic Modeling with 4 topics (LDA model) \\

\textbf{Topic \#0 : } class, file, example, data, public, type, classes, gt, lt, interface, client, objects, method, xml, database\\
\textbf{Topic \#1 : } project, version, new, development, released, web, application, release, server, framework, sun, apache, available, users, based\\
\textbf{Topic \#2 : } object, api, org, string, return, method, function, called, case, private, class, reference, add, added, net\\
\textbf{Topic \#3 : } java, code, language, source, following, implementation, platform, software, applications, oracle, provides, program, features, run, based\\

Our aim was to find 4 topics but the result was not satisfactory as the words similar to the top words for each topic was not good for classifying questions. However, we found that topics were created around the concepts of java eg. topics related database, networking etc.

\section{Word2Vec and Doc2Vec}

In the paper they did categorization of questions into four categories from where we discovered the the top word corresponding to the each category as follows :

\begin{center}
 \begin{tabular}{| c | c |} 
 \hline
 Question Category & Top Word found \\ [0.5ex] 
 \hline
 Debug & debug \\ 
 \hline
 Need To Know & explain \\
 \hline
 How to do it & implement \\
 \hline
 Seeking different solution & suggest \\ [1ex] 
 \hline
\end{tabular}
\end{center}

We need to set the context window which determines how many words before and after a given word would be included as context words of the given word.\\
We ran Word2Vec for context window size of 10. Then we found the most similar words from the model trained corresponding to the top words found above. Here also, as above WikiPedia articles were used.
The vectors obtained can be used to find a word similar to a given word, or to perform clustering.

\subsection{Results of Word2Vec}

4 topics( Word2Vec )  : Window Size = 10\\
The numerical value against each word indicate how similar it is to the top word of that category.\\

\textbf{suggest}: [('advise', 0.9350534081459045), ('explain', 0.8972327709197998), ('tell', 0.8767426013946533), ('help', 0.8588043451309204), ('assist', 0.8563621640205383), ('bear', 0.853961706161499), ('suggestion', 0.8530131578445435), ('know', 0.8440178632736206), ('enlighten', 0.8383816480636597), ('assistance', 0.8383312225341797)]\\

\textbf{debug}: [('isdebugenabled', 0.6271491646766663), ('stacktrace', 0.6043941378593445), ('getcontentlength', 0.6030460000038147), ('inflight', 0.5879602432250977), ('p\_get\_class\_schedule', 0.5865908265113831), ('refreshing', 0.5822206735610962), ('setdebug', 0.5703291893005371), ('defaulting', 0.5698220729827881), ('loglevel', 0.565199613571167), ('logger', 0.5500662422180176)]\\

\textbf{implement}: [(‘algorithm', 0.8106338977813721), ('abstract', 0.7818029522895813), ('serializable', 0.7369078993797302), ('interface', 0.7162636518478394), ('extend', 0.7146245241165161), ('implementing', 0.7100733518600464), ('extending', 0.691871166229248), ('subclass', 0.6590582132339478), ('webapplicationinitializer', 0.6560215353965759), ('class', 0.6359965801239014)]\\

\textbf{explain}: [('quite', 0.8827763795852661), ('clarify', 0.8689993023872375), ('perhaps', 0.8645872473716736), ('especially', 0.863167941570282), ('better', 0.8565487861633301), ('difficult', 0.8529151678085327), ('express', 0.8518675565719604), ('situation', 0.849044680595398), ('performance', 0.8483530879020691), ('fast', 0.8476318120956421)]\\

\subsection{Results of Doc2Vec}

The numerical value against each word indicate how similar it is to the top word of that category/topic.\\

\textbf{suggest}: [('advise', 0.7253733277320862), ('help', 0.7107036113739014), (‘guide', 0.7041284441947937), ('assist', 0.6895503997802734), ('tell', 0.6714411973953247), ('enlighten', 0.6282055377960205), ('know', 0.6244207620620728), (‘bear', 0.6184274554252625), ('excuse', 0.5978437066078186), ('suggestion', 0.5930529832839966)]\\

\textbf{implement}: [('implementing', 0.5342041850090027), (‘algorithm', 0.5311474800109863), (‘extend', 0.52436097860336304), (‘execute', 0.5210355854034424), (‘function', 0.5163625431060791), ('abstract', 0.5103716850280762), ('interface', 0.46266812086105347), ('public', 0.3308292627334595), ('keypressthread', 0.3260781168937683), ('instantiable', 0.3248624801635742)]\\

\textbf{debug}: [('isdebugenabled', 0.3359779119491577), (‘rectify’, 0.3264111280441284), ('stacktrace', 0.32063740491867065), ('logcat', 0.32000619173049927), ('setdebug', 0.3003194332122803), ('exec', 0.2963804602622986), ('logger', 0.2900662422180176), ('getanonymouslogger', 0.2807658016681671),  ('conn', 0.2756619453430176), ('rootlogger', 0.26862865686416626)]\\

\textbf{explain}: [(‘clarify', 0.5776931047439575), (‘difference', 0.5335918664932251), ('quite', 0.5324383974075317), (‘express', 0.5210134983062744), ('understanding', 0.4876328408718109), (‘sense', 0.48241132497787476), ('perhaps', 0.470624178647995), ('might', 0.4641563892364502), ('much', 0.4596584439277649), ('perhaps', 0.470624178647995), ('situation', 0.440624178647995)]\\

\section{Classifying}
We applied classifiers such as Naive Bayes, Logistic Regression and LSTM to predict the result. For the evaluation( of 4000 posts data ), we kept 80\%
data as training data i.e. 3200 samples and 20\% data for
testing i.e. 800 samples per class. After we got a list of words and the different words to which they are similar, we manually annotate the data according the class we want. The How-to-do-it category is very close to scenario in which a developer has a programming task at hand and need
to solve it. For this reason, in our approach, we only consider
Q\&A pairs that are classified as How-to-do-it.

\begin{center}
\captionof{table}{Accuracy of Classifiers Used}
\begin{tabular}{|c|c|c|}
	\hline
	Classifier & Accuracy\\
	\hline
    Naive Bayes & 76.184\%\\
    \hline
    Logistic Regression & 82.573\%\\
	\hline
     LSTM & 78.121\%\\
	\hline
\end{tabular}
\end{center}

\chapter{Approach}\label{chapter:Approach}

Our objective is to automatically tag each line of a given source code, in such a manner that the line matches it’s associated named entity.  To achieve this, we used:\\ (1) Entity Discovery, \\(2) Entity Profile Construction \\ (3) Entity Linking. \\

\section{Entity Discovery}
We extracted 20 entities from TutorialsPoint, and then applied this scheme to extract more. So, what we did was for each entity which appears in title of question, we extracted those questions. Then divided this data into 80\% and 20\%. For the 80\% data, we did POS tagging of questions, and then used those patterns which have a minimum support of 0.1 (normalized). For each pattern we saw if it existed in 20\% of data, and if yes we added it in vector. Now, we sorted all the generated patterns from maximum to minimum, and then took 5 first distinct entities. After discovering 5 entities from a pattern, we stopped to remove redundancy and avoid not so useful entities. \\
Consider example: 
For array, the most frequent pattern was NN IN DT ENTITY IN NNS where ENTITY is the placeholder for array. As an example, the SO title, "How to determine type of object/NN in/IN an/DT array/ENTITY of/IN objects/NNS" has this frequent pattern. Same pattern appears in another title, "Get an array of int/NN from/IN a/DT string/ENTITY of/IN numbers/NNS". So we gather that both array and string have the same PoS sequence. This is how we discovered other entities.\\
We could extract 2000( 200 of which are useful ) more useful entities after this. 

\section{Entity Profile Construction}

In this step, we create a profile for each entity that we had discovered, by matching them to their respective syntactical patters. 

If we are interested in the entity “conditional”, so “condition” would have syntactical structure composed of a few tokens, now this structure would be repeated across multiple codes which would be attached to the keyword “conditional”.

So we want to discover these patterns in source code that are associated with specific entities (like array or conditional). For array we see that it can be best matched with, [ ], whereas conditional could be best matched with, if ( ).

We need to identify the most appropriate n-grams that represent a specific entity from dataset. So we use the TF-IDF over n-grams to identify the syntactic patterns that are most associated with a given entity. To compute term frequency  \textit{tf} (t,g) of an n-gram g, we use the SO posts containing the entity name in title (Table 5.1). For IDF computation, we use all SO posts. Thus, we use the TF-IDF weight = tf (t, g) x log \textbar D\textbar / df(g) , where \textbar D\textbar - the total number of posts in SO and df(g) is the number of such posts containing the n-gram, g. Table 5.2 shows the results of these steps for a few entities.

\section{Results for Entity Profile Construction}

\begin{center}
\captionof{table}{Patterns and frequencies for \textit{\textbf{loop}} in Java snippets}
\begin{tabular}{|c|c|c|}
	\hline
	Uni-Gram Pattern & Normalized Frequency\\
	\hline
    log & 1.000\\
    \hline
    boolean & 0.985\\
	\hline
     for & 0.942\\
     \hline
     while & 0.840\\
	\hline
\end{tabular}
\quad
\begin{tabular}{|c|c|c|}
	\hline
	N-Gram Pattern & Normalized Frequency\\
	\hline
    for ( ; ; ) \{ & 1.000\\
    \hline
    for ( : ) ; & 0.858\\
	\hline
     ( + ( ) ) ; & 0.577\\
     \hline
     for ( ; ; ) ; & 0.566\\
	\hline
\end{tabular}

\end{center}

\begin{center}
\captionof{table}{Precision@4 along with top pattern discovered for some of the entities}
\begin{tabular}{|c|c|c|}
	\hline
	Entity & p@4 & Most Relevant Pattern\\
	\hline
    conditional & 1.00 & if () \{\\
    \hline
    array & 1.00 & [] \\
	\hline
     loop & 0.75 & for ( ;; )\\
     \hline
     increment & 0.75 & ++ \\
	\hline
    decrement & 0.75 & -- \\
    \hline
    parameter & 0.50 & ()\\
    \hline
\end{tabular}

\end{center}

\section{Entity Linking}
In this step, we annotate every line of a given source code, by the entity names that appear in that line. 

Every line of code is cleaned by removing user defined terms, to just focus on programming keywords. After the line is read and cleaned, we start with treating each term being a uni-gram and then switch to bi-grams, tri-grams, and so on, until all the n-grams are covered.

We used Entity-Profile created as once n-grams are created of a line, we match those with the syntactical patterns of that entity (from Entity-Profile created) and determine if they match.

Once an entity has been determined for a line of code, we annotates that line with the entity name as a comment. This further helps in searching within a source code by using regular keywords. A given line can have multiple entities, however we need to mark the most relevant one. Therefore to keep it concise, we have annotated no more than four entities per line.

\section{Results of Entity Linking}

\begin{center}
\includegraphics[width=\textwidth]{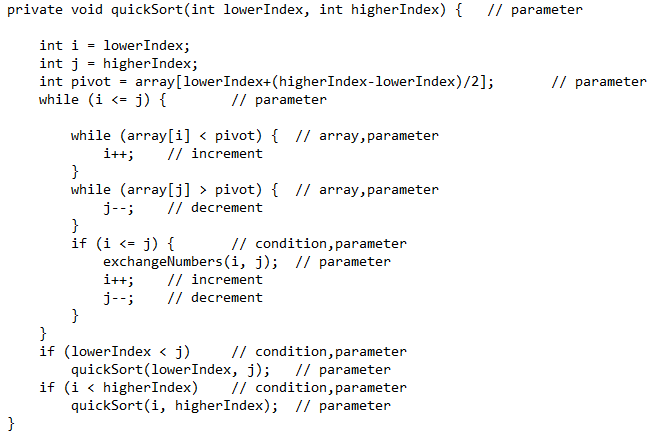}
\textbf{Fig: Annotated Code after Step 3}
\end{center}

\section{Performing Search}
We used Apache Lucene for this. Lucene takes all the documents, splits them into words, and then builds an index for each word. The index contains word id, number of docs where the word is present, and the position of the word in those documents. So when given a single word query, it just searches the index and returns the result. For multi-word query just take the intersection of the set of files where the words are present.

\chapter{References}\label{chapter:References}
\begin{enumerate}
\item \url{https://radimrehurek.com/gensim/models/word2vec.html}
\item \url{https://radimrehurek.com/gensim/models/doc2vec.html}
\item \url{https://www.analyticsvidhya.com/blog/2016/08/beginners-guide-to-topic-modelingin-python/}
\item \url{https://en.wikipedia.org/wiki/Wikipedia:WikiProject_Java/List_of_articles}\item \url{https://en.wikipedia.org/wiki/Wikipedia:WikiProject_Java/List_of_articles}
\item \url{http://dl.acm.org/citation.cfm?id=3018691}
\item \url{http://www.facom.ufu.br/~marcmaia/ICPC2014preprint.pdf}
\item \url{https://pdfs.semanticscholar.org/90af/f07df953e95dd6982b91462c64ef38f53fdc.pdf}
\end{enumerate}

\end{document}